
\input phyzzx
\sequentialequations
\overfullrule=0pt
\tolerance=5000
\nopubblock
\twelvepoint

\REF\bombelli{L. Bombelli, R. Koul, J. Lee, and R. Sorkin,
Phys. Rev.  {\bf D34} (1986) 373.}

\REF\thesis{C. Holzhey, {\it Princeton University thesis}, 1992,
(unpublished)}

\REF\srednickione{M. Srednicki,
Phys. Rev. Lett. {\bf 71}, (1993) 666 }

\REF\ccfw{C.G.Callan and F. Wilczek,
{\it On Geometric Entropy},
hep-th-9401072}

\REF\dowker{J.S.Dowker,
{\it Remarks on Geometric Entropy},
MUTP-94-2,hep-th-9401159}

\REF\kabat{D. Kabat and M.J.Strassler,
{\it A Comment on Entropy and Area},
RU-94-10,hep-th/9401125}

\REF\susskindfour{L. Susskind,
{\it Some Speculations About Black Hole Entropy in String Theory},
RU-93-44,hep-th/9309145}

\REF\hlw{ C. Holzhey, F. Larsen and F. Wilczek,
{\it Geometric and Renormalized Entropy},
PUPT-1454, IASSNS-93/88, hep-th/9403108}

\REF\susskindfive{L. Susskind and J. Uglum,
{\it Black Hole Entropy in Canonical Quantum Gravity and Superstring Theory},
SU-ITP-94-1, hep-th/9401070}

\REF\unruh{W. G. Unruh,
Phys. Rev. {\bf D14} (1977) 870}


\REF\cardy{J. L. Cardy,
{\it Conformal Invariance and Statistical Mechanics},
in {\it Fields, Strings, and Critical Phenomena}
Les Houches, Session XLIX, 1988, eds. E.Br\'{e}zin and J. Zinn-Justin.}

\REF\faddeev{L. Faddeev and Slavnov,
{\it Gauge Fields, Introduction to Quantum Theory},
Benjamin/Cummings (1980)}

\REF\itzykson{C. Itzykson and J. B. Zuber,
{\it Quantum Field Theory},
McGraw Hill (1980)}

\REF\wavelet{M. B. Ruskai {\it et al.},
{\it Wavelets and Their Application},
Jones and Bartlett (1992)}

\REF\soper{D. E. Soper,
Phys. Rev. {\bf D18}(1978) 4590}

\def\fint{\int {d\omega\over 2\pi}}

\line{\hfill PUPT 1480, IASSNS 94/51}
\line{\hfill hep-th/9408XXXX}
\line{\hfill August 1994}

\titlepage
\title{Geometric Entropy, Wave Functionals, and Fermions}

\author{Finn Larsen\foot{Research supported in part by Danish National
Science Foundation Fellowship.\ \ \ larsen@puhep1.princeton.edu}}
\centerline{{\it Department of Physics }}
\centerline{{\it Joseph Henry Laboratories }}
\centerline{{\it Princeton University }}
\centerline{{\it Princeton, N.J. 08544 }}
\vskip .2cm
\author{Frank Wilczek\foot{Research supported in part by DOE grant
DE-FG02-90ER40542.~~~wilczek@iassns.bitnet}}
\vskip.2cm
\centerline{{\it School of Natural Sciences}}
\centerline{{\it Institute for Advanced Study}}
\centerline{{\it Olden Lane}}
\centerline{{\it Princeton, N.J. 08540}}
\endpage

\abstract{We develop techniques for calculating the ground state
wave functional and the geometric entropy for some simple field
theories.  Special attention is devoted to fermions, which present
special technical difficulties in this regard.  Explicit calculations
are carried through for free mass bosons and fermions in two
dimensions, using an adaptation of Unruh's technique to treat
black hole radiance.}

\endpage

\chapter{Introduction}

Recently there has been great interest in the concept of geometric
entropy [\bombelli-\susskindfive].
Almost all of the explicit
work has been done for free scalar bosons, although one can adapt
conformal field theory arguments to cover some additional
special cases.
Among several possible generalizations, perhaps the most interesting
is to fermions.  Fermions play a role in several
of the speculations
regarding geometric entropy, especially those involving
its analogues in
superstring theory.
The ground state wave functional and the path integral,
fundamental objects from any perspective and
certainly central to any discussion
of geometric entropy, have quite a different character for fermions than
for bosons.  Whereas for bosons the states of the quantum field
theory can be labelled by classical field configurations, since the field
operators on a complete hypersurface constitute a complete set of commuting
observables, the corresponding structure for fermions is less transparent.

Let us recall the basic construction used to define geometric entropy.
One divides space into two regions, which
we shall call inside and outside.
We consider a quantum
field theory defined in all of space, and possessing a
complete set of local observables $\xi(x)$. We divide these
into two disjoint subsets according to whether their argument
$x$ lies on
the inside or on the outside.
Any state may be labelled by its amplitude as a function of the
the eigenvalues $\xi_{\rm in}, \xi_{\rm out}$ of the $\xi(x)$
with $x$ lying on the inside or the outside respectively.
Now consider a
definite pure state described by the
wave functional $\Psi[\xi_{\rm in},\xi_{\rm out}]$.
In the body of this paper, we shall be concerned with the ground state
exclusively, and when we speak of geometric entropy without further
specification we shall have this
case in mind.   Then the
density
matrix appropriate to an observer who only has access to the
outside region is obtained by tracing over
the variables localized in the
in-region:
$$
\rho(\xi_{\rm out}^{1},\xi_{\rm out}^{2})=\sum_{\xi_{in}}
{\bar\Psi}[\xi_{in},\xi_{out}^{1}]\Psi[\xi_{in},\xi_{out}^{2}]~.
\eqn\trace
$$
This density matrix in general
no longer describes a pure state, and one
defines the geometric entropy relative to the given state and the
given division of space as
$$
S_{\rm geom} = - {\rm Tr} \rho {\rm ln} \rho  ~.
\eqn\entropy
$$
This $S$ is a measure of the strength of the
correlations between the inside
region and the outside region, information concerning which is
lost in the process of tracing over
the inside region.   It has several qualitative
properties that are rather different from those of thermodynamic
entropy; notably it is not extensive, and its value is invariant
under
interchange of ``inside'' with ``outside''.
In field theories with sufficiently singular ultraviolet behavior
(that is, in
essentially all ordinary -- non-topological -- field theories)
the geometric entropy is dominated by short-range correlations.
In the
ultraviolet limit all theories look scale invariant, and
furthermore the important correlations arise from field
fluctuations in the direction of the normal at the boundary.
As a matter of technical convenience, rather than conceptual necessity
we shall, below, often use free massless fields in 1+1 dimensions as
our working material.  Due to the considerations just mentioned, this
specialization is less severe than might appear at first sight.
(In this regard, the interested reader might like to refer to
Equation 36 of  [\ccfw ] as an illustration of the special significance
of the 1+1 dimensional massless theories.
Such
theories, of course, also occupy center stage in superstring theory.)

Although we will not pursue it much further in this paper,
let us briefly note that one can also define the geometric entropy
for mixed states, \eg\ thermal states.  In that context several of
the points made above require modification.  In particular, there is
no longer symmetry between inside and outside; futhermore, the
{\it difference\/} between the thermal geometric entropies at
different temperatures is not expected to be dominated by ultraviolet
behavior, and it should be approximately extensive for large volumes -- the
bulk geometric entropy density coincides with the bulk thermodynamic
entropy.

We have considered the geometric entropy in general
conformal field theories previously,
and obtained quantitative results which coincide with the
ones presented here [\hlw].
The goal of the present work is, in a sense, to do it again the
hard way --  working with
field variables (especially, fermion field variables )
directly, without invoking conformal symmetry.
Our goal is, by being very pedestrian and explicit,
to sharpen our understanding of the technical issues involved,
particularly in defining the wave functionals and the path integrals.

As an example of the sort of subtlety that arises, consider the following.
In their standard formulation, relativistic fermions obey a linear equation,
arising from a quadratic action.  One is accustomed to evaluating quadratic
path integrals by substituting the solutions of the classical
equations of motion (with the appropriate boundary conditions)
back into the action.  But here the action will simply vanish, and
by proceeding naively one
obtains a trivial --  wrong -- answer.

Below, by exercising more care, we shall be able to obtain meaningful,
useful expressions in this and similar contexts.
Having these in hand,
we will proceed to evaluate the density matrix and geometric
entropy for free massless bosons and fermions in 1+1 dimensions.
When the dust
settles,  we are left with a very convenient
formalism that
treats bosons and fermions on an equal footing.
Here, as in most other work on geometric entropy, the
insight that allows us to pursue matters to the very end is an inspired
choice of variables, first used
by Unruh [\unruh],
who used it in a different but related context.
But there are some small surprises in the details:
our generalization of the
Unruh {\it ansatz\/} contains a funny coefficient, signs must be treated
very carefully, and boundary conditions are nontrivial.

\chapter{The Wave Functional}

To discuss geometric entropy we need to
describe our reference state in terms of local variables.
In ordinary quantum mechanics we can
project
on eigenstates $|q\rangle$ of the position operator
$\hat q$, thus defining
$\Psi(q,t) = \langle q|\Psi(t) \rangle$.
For
field theory, we want to
choose a complete set of local variables generalizing
$\hat q$. The field operators $\hat\phi(x)$ form
a natural choice,
but it is by no means unique.  We might, for example, choose
instead to diagonalize the canonical momenta $\hat{\pi}_\phi (x)$.
It is by no means obvious at first sight what
constitutes
the ``natural'' choice is for fermions;
we shall
see it is in a sense a linear combination of the two just mentioned.

\section{Schr\"{o}dinger Equation and Path Integral}

Although the definition of geometric entropy does not
refer explicitly to dynamics, we will find it useful to
consider solutions to the time-dependent Schr\"{o}dinger
equation
$$
 i {d\over dt} |\Psi\rangle  = {\hat H}|\Psi\rangle
\eqn\schreqn
$$
where $\hat {H}$ is the Hamiltonian operator.
The formal solution is
$$
|\Psi(t_f)\rangle = Te^{-i{\hat H}(t_f-t_i)} |\Psi(t_i)\rangle
\eqn\formal
$$
where T denotes time-ordering.
Time-ordering is an essential requirement, because the
Hamiltonian is an integral over
objects that do not mutually commute.

The formal expression \formal\ for the wave function
is difficult to evaluate
explicitly for specific field theories,
and the time-ordering
obscures the Lorentz invariance of the theory.
Indeed, the Hamiltonian as well as
the time-ordering procedure depend explicitly on the choice
of reference frame.  Fortunately, one has the identity
$$
\Psi[\phi_f] = \langle\phi_f| T e^{-i{\hat H}
(t_f-t_i)} |\phi_i \rangle
= \int {\cal D}\phi~ e^{iA(\phi)}~.
\eqn\lh
$$
In this path integral, one integrates
over all field configurations for which
the field is a prescribed value $\phi_i$ at some early time $t_i$ and
equal to $\phi_f$ at the final time $t_f$. The field configuration
at the early time selects a specific state. The measure will
be constructed below. If the action $A(\phi)$
and the measure ${\cal D}\phi$ are Lorentz invariant, the transformation
law for the wave functional is simply
given by appropriate change in boundary conditions.


\section{Symmetry and the Geometric Derivation}

We recall a recent simple derivation [\cardy ] of
\lh , that is very instructive for our purposes.
The Hamiltonian expression satisfies a simple first order
differential equation.
Since both expressions reduce to 1
for $t_f = t_i$,
to prove their identity it suffices to show that the
Lagrangian expression satisfies the same differential equation.
Thus we must evaluate
$$
\delta\int{\cal D}\phi~e^{i\int_{t_i}^{t_f} {\cal L}(\phi)~d^d x}
$$
where the variation is a change in $t_f$ to $t_f+\delta t_f$.

The definition of the energy momentum tensor $T_{\mu\nu}$ is
$$
dA = - {1\over 4\pi} \int T_{\mu\nu} dg^{\mu\nu} d^d x
\eqn\actionvar
$$
under a change $dg^{\mu\nu}$ in metric.
The variation $\delta$
may be implemented by choosing
a new $g_{00}=(1+2{\delta t_f\over t_f-t_i}+{\rm const})$ .
Following [\cardy] (using, in essence, a special
case of the Schwinger action principle),
we require the expression to be invariant under
changes in coordinates, thus deriving
$$
\delta
\int{\cal D}\phi~e^{iA} \simeq {i\over 4\pi}
\int {\cal D}\phi e^{iA}\int T_{\mu\nu}dg^{\mu\nu} d^d x
\eqn\zebra
$$
for true changes in geometry. Notice the sign of this expression.
Next, we can use conservation of (integrated!)
energy-momentum to put this
in the form
$$
{\delta\int {\cal D}\phi~e^{iA}\over \delta t_f} =
- {i\over 2\pi} {1\over t_f - t_i} \int {\cal D}\phi
(\int T_{00} d^d x)~e^{iA}= -i{\hat H}\Psi[\phi]~,
\eqn\yak
$$
which establishes \lh .
The action  of Hamiltonian operator $\hat H$ on the wave functional
is defined by the last equation, {\it i.e}. it is
found by evaluating the operation occurring in \yak\ on
the
final slice of the functional integral.

For the preceding derivation, the crucial property
of the path integral
measure employed is invariance under time dilation.
We will also require invariance under translation with a function.
A measure satisfying these properties will
be constructed in the next sections, and will involve some
surprising subtleties.

To motivate the detailed
investigation, let us elaborate further upon an difficulty
with
with the na\"{\i}ve application of the path integral for fermions,
that we already touched on briefly.
For simplicity, we consider a Weyl
fermion in two dimensions. On shell this
can be represented as a function of (say) $x+t$, or in
complex variables as a holomorphic function of $z$.
The action is
$A={1\over 2\pi}\int\psi{\bar\partial}\psi$.
Now it appears that, at the saddle point,
this action vanishes, since the equation of motion is simply
${\bar\partial}\psi=0$.  One the other hand,
consider the Hamiltonian
expression for the wave functional: the Hamiltonian
$\hat{H}={i\over 4\pi}\int\psi\partial_\sigma\psi$
certainly
does not vanish for holomorphic $\psi$s.
How, then, can the two
expressions \lh\ for the wave functional
possibly agree?

To address this question recall that
in the derivation of the path integral formula for the wave function
we exploited the feature that
the measure is invariant under a rescaling
of the time variable.
But under such a rescaling, $z$ mixes with $\bar z$. It is therefore
important that, as intermediate configurations in the path integral,
we use a set of states that is closed under time translations.
That means that even for chiral particles such as
Weyl fermions,
one cannot impose
the equation of motion on intermediate states.
To get a simple path integral we must
allow both right and left
movers in intermediate states, even for a Weyl particle.
Once we do that, it becomes less mysterious why
the action does not necessarily vanish.
Note that in this reasoning
we tacitly assumed that a stationary phase method is applicable
and exact for the quadratic action, or in other words that
by imposing the equations of motions found by varying the action
with respect to the field, we find a field $\psi_{\rm cl}$ that
saturates the functional integral.
In the next sections we will construct the
path integral that has this desirable feature
explicitly, and verify that indeed such a classical
field exists even for chiral fermions --
but the classical field will include
both left and right movers.

\section{The Holomorphic Path Integral}

A popular modern way to construct path integrals uses the
holomorphic representation, also called
the Bargmann-Fock representation [\faddeev--\itzykson].
In this representation we work in bases
that diagonalize the positive and negative frequency parts
of the field, rather than the
Hermitean operators ${\hat\phi}$ and
${\hat\pi}\propto\partial_t{\hat\phi}$ which are perhaps
more natural from
an intuitive
standpoint.  In the holomorphic representation we diagonalize
the $\hat{\phi} (x) + i\hat{\pi}(x)$
and thus  also their Fourier transforms,
denoted $\hat{\phi}_k$.
It is in terms of the these latter
operators that the free Hamiltonian can be expressed as a sum of
fundamental harmonic oscillators, i.e. ${\hat H}=\int{dk\over 2\pi}
\omega_k {\hat\phi}_k^\dagger {\hat\phi}_k$.
In principle all we need for the calculation
of the geometric entropy (or anything else)
is the ability to evaluate a
complete set of commuting local variables,
so either choice will serve.

For a boson field $\phi$ we use
the coherent ket states
$$
|\phi\rangle
=\prod_k e^{-{1\over 2}{\bar\phi}_k\phi_k}e^{{\hat\phi}_k^\dagger \phi_k}
|0\rangle~.
\eqn\cohstate
$$
This state is an eigenfunction of the positive frequency part of
the field operator, {\it i.e}.
${\hat\phi}_k|\phi\rangle=\phi_k|\phi_k\rangle$. The index
$k$ labels the degrees of freedom which, for a free
scalar field, can be identified with the momentum.
For massless fields
in two dimensions $k$ may be restricted
to take either only positive or only
negative values, if one of these restrictions is imposed the field is chiral.
The bras are obtained by complex
conjugation of the kets.  They diagonalize the negative
frequency part of the field to the acting to the left:
$\langle\phi|{\hat\phi}_k^\dagger
=\langle\phi |{\bar\phi}_k$.  There is a completeness relation
$$
\int \prod_k d{\bar\phi}_k d\phi_k ~|\phi\rangle\langle\phi| = 1~,
\eqn\compl
$$
which implicitly normalizes the measure.

Now we can evaluate the evolution operator, as follows:
$$
\eqalign{
U({\bar\phi}_f | \phi_i ) &\equiv
\langle{\bar\phi}_f|Te^{-i\hat{H}t}|\phi_i\rangle \cr
&= \int\prod_k (\prod_{j=1}^{n-1}d{\bar\phi_j}d\phi_j)
\prod_{j=0}^{n-1}\langle{\bar\phi}_{j+1}|
e^{-i \omega_k {\hat\phi}_k^\dagger {\hat\phi}_k
\Delta t_j}|\phi_j\rangle \cr
&= \int\prod_k(\prod_{j=1}^{n-1}d{\bar\phi_j}d\phi_j)
\prod_{j=0}^{n-1}\langle{\bar\phi}_{j+1}|e^{-i\omega_k\Delta t_j}
\phi_j\rangle \cr
&= \int\prod_k (\prod_{j=1}^{n-1}d{\bar\phi_j}d\phi_j )
\prod_{j=0}^{n-1}
e^{-[{1\over 2}({\bar\phi}_{j+1}{\phi_{j+1}-\phi_j\over\Delta t_j}-
{{\bar\phi}_{j+1}-{\bar\phi}_j\over\Delta t_j}\phi_j)
+i\omega_k {\bar\phi}_{j+1} \phi_j)]\Delta t_j }~.\cr }
\eqn\evop
$$
In this manipulation,
we handled the time ordering by inserting a sequence
of complete sets and
used the fact
that coherent states remain coherent under time evolution.
To avoid
proliferation of indices, we suppressed indices $k$ on $\phi$ and
${\bar\phi}$; the remaining indices
label the time slices,
with the understanding that times $0$ and $n$
replaces $i$ and $f$. Taking $\Delta t_j=t_{j+1}-t_j
\rightarrow 0$ and using the notation
${\dot\phi}={\phi_{j+1}-\phi_j\over\Delta t_j}$ we find
$$
U( {\bar\phi}_f | \phi_i )=\int {\cal D}{\phi}
{}~[e^{{1\over 2}({\bar\phi}_f\phi_f+{\bar\phi}_i\phi_i )}
e^{-\int dt~[{1\over 2}({\bar\phi}{\dot\phi}-{\dot{\bar\phi}}\phi )
+i\omega {\bar\phi}\phi ]} ]
e^{-{1\over 2}({\bar\phi}_f\phi_f+{\bar\phi}_i\phi_i )}~
\eqn\holint
$$
in which, of course, we have introduced a new notation for the measure.
The purpose of the additional, vacuous exponentials will be explained
shortly.

A Gaussian integral worthy of the name should be
calculated exactly using the stationary phase method.
Here a first cut at such an evaluation would involve imposing
the equations of motion found by varying ${\bar\phi}$ and $\phi$
separately\foot{We
ignore the determinant, which does not depend on the boundary
conditions}.
However to obtain
a precise prescription the correct measure must be
kept in mind. That measure included only variables with indices $1$
through $n-1$, so in applying the stationary phase method we must
not allow variations at the end points.  In the given variables
this is an awkward constraint,
because as we vary ${\bar\phi}$ with ${\bar\phi}_f$ fixed
we inevitably vary ${\bar\phi}_i$, and similarly for
the holomorphic variables.
By adjusting the boundary term appropriately,
one is able to organize the expression as in
\holint ,
where the expression in the
square brackets allows the full variation.
Indeed, upon translating back to the
discrete notation, one easily checks that
the bracketed expression
contains neither $\phi_f$ nor ${\bar\phi}_i$.
Evidently the symbols with those
names should be considered functions of the other boundary conditions,
not as independent parts of the path integral.
By way of contrast, in
the factor outside the brackets, the bar really does denote
complex conjugation.

The expression in the square brackets is the standard expression
for the evolution operator in the holomorphic representation.
Finding and integrating the equations of motion, one readily
calculates
$$
U({\bar\phi}_f|\phi_i)=
\prod_k~[e^{{\bar\phi}_f~e^{-i\omega_k (t_f-t_i)}\phi_i}]
e^{-{1\over 2}({\bar\phi}_f\phi_f+{\bar\phi}_i\phi_i)}
\eqn\bfevol
$$
This expression could also be derived directly, using simple
properties
of coherent states.

Taking the initial state to be vacuum, {\it i.e}.
$\phi_i=0$, we find the
vacuum wave functional to be a simple normalized Gaussian.
This final result should
come as no surprise, because the Hamiltonian
is zero acting on the vacuum,
so $U({\bar\phi}_f |0)=\langle{\bar\phi_f} |0\rangle$,
leaving only
the normalization of the coherent state.
This satisfying result exhibits also, unfortunately, the
essential triviality of the procedure: the insertion of intermediate
states, and subsequent integration, is vacuous.  The entire calculation
effectively occurs at one time slice, and one does not find the
Lorentz or conformal symmetries exhibited explicitly.

\section{Feynman's Path Integral}

To exhibit these symmetries,
we would like to find a form of the path integral
that, written in real space, resembles the one derived from
geometrical considerations.  For this purpose, we return to the
original Feynman construction of the path integral.
The Hamiltonian is
$$
{\hat H} = \int {dk\over 2\pi} \omega_k{\hat\phi}_k^\dagger {\hat\phi}_k =
\int {dk\over 2\pi}
{1\over 2\omega_k}({\hat p}_k^2+\omega_k^2 {\hat q}_k^2);~~~
\eqn\ham
$$
where
${\hat\phi}_k={\omega_k{\hat q}_k+i{\hat p}_k\over\sqrt{2\omega_k}}$ and
${\hat\phi}_k^\dagger={\omega_k{\hat q}_k-i{\hat p}_k\over\sqrt{2\omega_k}}$,
so that $[{\hat\phi}_k,{\hat\phi}_k^\dagger]=2\pi\delta(k-k^\prime)$.
As a basis in the Hilbert space we introduce the overcomplete
set $|q\rangle $ with the property ${\hat q}_k|q\rangle = q_k|q\rangle$.
 From the commutation relations
$[{\hat p}_k,{\hat q}_k]=-2\pi i\delta(k-k^\prime)$ we derive
that ${\hat p}_k$ act on such states as ${\hat p}_k|q\rangle = i{d\over dq_k}
|q\rangle$. Introducing also eigenstates of the momentum operator
we find $\langle p|q\rangle = e^{-ip_k q_k}$. The operators ${\hat q}_k$ and
${\hat p}_k$ are Hermitean so their eigenvalues are real.
 From the Hamiltonian we see that vacuum is defined by ${\hat\phi}_k|0\rangle
=0$,
and integrate to find
$$
\Psi[q] =\langle q|0\rangle \propto \prod_k e^{-{\omega_k\over 2} q_k^2}
\eqn\grdstate
$$

Normalizing the measure so $\int dq_k~ e^{-\omega q_k^2}\equiv 1$,
$$
\int dq~|q\rangle \langle q| = 1 = \int dp~|p\rangle\langle p|
\eqn\norm
$$
(we omit the indices and denote the numbers $q_k$
collectively as $q$, and similarly for $p_k$),
we are ready to derive the path integral:
$$
\eqalign{
U(q_f|q_i) &\equiv \langle q_f| ~Te^{-i{\hat H}t}|q_i\rangle \cr
&=\int\prod_{j=1}^{n-1}dq_j
\prod_{j=0}^{n-1} dp_j \langle q_{j+1}|p_j\rangle\langle p_j|e^{-i{\hat H}
\Delta t_j}|q_j\rangle \cr
&=\int\prod_{j=1}^{n-1}dq_j \prod_{j=0}^{n-1}dp_j~e^{ip_j (q_{j+1}-q_j)}
e^{-{i\over 2}(p_j^2+\omega_k^2 q^2_j)\Delta t_j} \cr
&\equiv \int {\cal D}q
\prod_k~e^{i\int dt {1\over 2}({\dot q_k}^2 -\omega^2_k q_k^2)} \cr
&=\int{\cal D}\phi~e^{i\int dt {1\over 2}
({\dot\phi}^2-({\nabla\phi})^2)} \cr}
\eqn\bpathint
$$
In this expression, we have
defined $t_n=t_f$, $t_0=t_i$, and $\Delta t_j=t_{j+1}-t_j$, and
used
a continuum notation.

We have gone into some detail here so that the main point comes out
clearly and unambiguously, as follows.
By construction, the variables $q_k$ are real.
This property must hold even it the field
is chiral, that is if the $k$ are
restricted to be positive (or negative). The
classical field includes both components nevertheless,
because that is the only way
it can be real.
All this conforms with our earlier remarks on the geometric
derivation of the path integral.
We realised in that context
that
one must allow both left and right movers
in the intermediate
configurations, so as to work
in a set that is closed under dilation of time.

To evaluate the integral, we write $\phi=\phi_{\rm cl}+\delta\phi$
where $\phi_{cl}$ is a solution of the equation of motion with the
specified
boundary conditions, while $\delta\phi$ vanishes on the
boundary but is otherwise arbitrary.
Inserting this decomposition in the
path integral, one easily sees that the cross terms disappear.
Hence the path integral factorizes, with only one factor depending
on the boundary conditions.  And that factor
is simply the integrand in
the path integral, {\it i.e}. exponential of
the classical action,
evaluated for the field $\phi_{cl}$.

It is easy to calculate the classical field even
for general boundary conditions, i.e. $\phi$ fixed
to be $\phi_f$ at $t_f$ and $\phi_i$ at $t_i$.
We are especially interested in the special case
where the initial state is the vacuum. This case
is most easily handled
by noticing that taking $t_f-t_i\rightarrow -i\infty$ the Hamiltonian
projects on to the ground state.  Indeed, finding the classical
field and taking this limit, with $\phi_i$
fixed at some finite value, has the same effect as removing the
components that approach $\infty$ in the limit, or alternatively
requiring that the classical field satisfies
$\phi\rightarrow 0$ in this limit. Hence the wave functional for
the vacuum state is calculated simply by imposing $\phi\rightarrow 0$
as $t\rightarrow -i\infty$ on the classical field.

To tie up this discussion let us relate this careful
explicit evaluation,
with an identified
measure $\prod_{k} dq_k$
and a definite
prescription for calculation of the path integral, to
the heuristic \lh . We shall illustrate this by reference
to the wave functional for a chiral field in 1+1 dimensional
Euclidean space.   We
find the classical field that satisfy the equations of motion and the
appropriate boundary conditions, and calculate the path integral for that
one configuration.
The vacuum state is identified
by the boundary condition $\phi\rightarrow 0$ as $t\rightarrow -i\infty$
or,
in Euclidean space, $\phi\rightarrow 0$ as $\tau\rightarrow -\infty$.
Imposing the reality condition, we write
$$
\phi_{\rm cl} = \int^\infty_{-\infty}
{dk\over\sqrt{4\pi|k|}}~[\phi_k~
e^{kz}+{\bar\phi}_k e^{k{\bar z}}]~.
\eqn\classfield
$$
Now we select the vacuum by restricting the integral to positive k.
Calculating the action for this field configuration, we find
$$
A_{\rm cl}(\phi_{\rm cl})=
{1\over 2\pi}\int
\partial\phi_{\rm cl}{\bar\partial}\phi_{\rm cl}=
{1\over 2} \int_0^\infty {dk\over 2\pi}
{\bar\phi}_k \phi_k
\eqn\classact
$$
The action integral was over the $\tau<0$ half plane.
Finally we write the wave functional
$$
\Psi[\phi] = \prod_k e^{-{1\over 2}{\bar\phi}_k\phi_k}~,
\eqn\xylophone
$$
a result that coincides
with the one found above in other ways.

\section{The Path Integral for Fermions}

With these experiences in mind, let us turn to
the question of finding a useful
prescription for the fermion path integral.
For free fermions the
Hamiltonian is again of the form
${\hat H}=\omega a^\dagger a$, but the creation and
annihilation operators satisfy anticommutation relations
$\{ a,a^\dagger\} =1 $. Introducing hermitean variables
${\hat q}={ a+a^\dagger\over\sqrt{2}}$ and
${\hat p}={ i(a^\dagger-a)\over\sqrt{2}}$, we find
$\{ {\hat p},{\hat q} \}=0$.
This relation does not allow for a realization
wherein
${\hat q}$ is diagonal and ${\hat p}$ is expressed as
a derivative operator. That would
require a nontrivial (anti)commutation relation.
Thus the road leading to the Feynman path integral appears
to be closed, and we must fall back on a version of the
holomorphic integral.
Indeed, one
can introduce a basis that diagonalize $a$ and represent $a^\dagger$
as an anticommuting derivative.
This leads us to the holomorphic representation
of the path integral,
with the additional subtlety that the holomorphic variables are
anticommuting, so that in manipulating the expressions we must take
care of their order.  In the bosonic (second-order) case this
form of the path integral was rather trivial and unsatisfying
geometrically, but as we shall now demonstrate the situation is
quite different in the fermionic (first-order) case.

As usual, we consider a massless field in two dimensions.
Introducing coherent states
$$
|\psi\rangle=\prod_k e^{-{1\over 2}{\bar\psi}_k\psi_k}e
^{{\hat\psi}_k^\dagger\psi_k}
|0\rangle
\eqn\fcohstate
$$
we preserve the
normalization of the state $\langle\psi|\psi\rangle$
and the resolution of the identity
$$
\int \prod_k d{\bar\psi}_k d\psi_k ~|\psi\rangle\langle\psi| = 1~.
\eqn\residen
$$
Since the Hamiltonian remains $\hat H=\int {dk\over 2\pi}
|k|{\hat\psi}_k^\dagger {\hat\psi}_k$
we find as before
$$
\eqalign{
U({\bar\psi}_f|\psi_i)
&= \int\prod_k (\prod_{j=1}^{n-1}d{\bar\psi_j}d\psi_j )
e^{-[{1\over 2}({\bar\psi}_{j+1}{\psi_{j+1}-\psi_j\over\Delta t_j}-
{{\bar\psi}_{j+1}-{\bar\psi}_j\over\Delta t_j}\psi_j)
+i|k| {\bar\psi}_{j+1} \psi_j)]\Delta t_j}\cr
&\equiv \int{\cal D}\psi \prod_k
e^{\int dt{1\over 2}
[{\dot{\bar\psi}}\psi-{\bar\psi}{\dot\psi}]+i|k|{\bar\psi}\psi} \cr
&\equiv \int {\cal D}\psi~e^{-{1\over 4\pi}
\int dt d\sigma ~\psi (\partial_t-\partial_\sigma)\psi }~.\cr}
\eqn\ftransamp
$$
In
transforming back to real space we have introduced
$$
\psi = \int {dk\over 2\pi}[ \psi_k e^{-ik\sigma}+{\bar\psi_k} e^{ik\sigma}]
\eqn\realpsi
$$
which by construction is real. We see that for fermions, unlike
for bosons, the holomorphic path integral leads to the invariant form
that we know from the geometrical derivation.  However here as
before these expressions are a little subtle.
In particular, in deriving and interpreting them
one may be tempted to give up the reality condition on the
field, in which case one must take boundary terms carefully into account
in evaluating them, and a trivial integral is found.

To evaluate the path integral without giving up the reality condition
we proceed as for bosons and write
the most general field $\psi=\psi_{\rm cl} +\delta\psi$, where $\psi_{\rm cl}$
satisfies the boundary conditions on $\psi$ while $\delta\psi$ is 0 on
the boundaries, that is the initial and final time slices.
Finally, choose
$\psi_{\rm cl}$ to be the unique real field that satisfies the
boundary conditions, as well as the Klein-Gordon ({\it not\/} the Weyl-Dirac)
equation.  (Any field with the proper boundary conditions might have been
used; the Klein-Gordon
equation is imposed only to insure uniqueness!).
Now note
that in
$$
{1\over 4\pi}\int \psi(\partial_t-\partial_\sigma)\psi
={1\over 4\pi}\int \psi_{\rm cl}(\partial_t-\partial_\sigma)\psi_{\rm cl}+
{1\over 4\pi}\int \delta\psi(\partial_t-\partial_\sigma)\delta\psi
\eqn\wombat
$$
mixed terms do not enter, due to the reality condition on
the field and the boundary
conditions on $\delta\psi$. The path integral measure ${\cal D}\psi$ is
invariant under translations with a fixed function $\psi_{\rm cl}$,
so we might as well take it to be ${\cal D}\delta\psi$. Then the
path integral factorizes. The fluctuation part is independent of the
boundary conditions and can be omitted.
We find that the path integral can be calculated by taking the action
of the field configuration $\psi_{\rm cl}$, as advertized.
At no point in the present derivation
did we use a variational principle -- specifically,
we never found that
we should impose the equation of motion on the classical field.

We have found that the fermionic path integral is naturally expressed using
real classical fermion fields to parametrize the initial and final states,
thus putting them on an equal footing. The interpretation of such an amplitude
between real states is not {\it a priori} clear, because the
original problem did not have real fields. It is implicit
in our derivation that the path integral expresses an amplitude
between coherent states. To find the ket-coherent state that
corresponds to a given real wave function, take the
spatial Fourier transform. The components with positive
wave vector provides the eigenvalues of the elementary oscillators
of the Weyl field. Conversely, the real classical field corresponding to given
eigenvalues of the elementary oscillators is found by using the eigenvalues
as Fourier components with positive k-vectors, and their complex conjugates
for the negative k-vectors. Thus the amplitudes that are expressed by
our unorthodox path integral are exactly the same as those that
are expressed by the more conventional holomorphic path integral.

Just as for bosons, on taking $t_i\rightarrow -i\infty$ with
$\psi_i$ fixed we find that $\psi_i$ disappears from the problem
and is replaced by the requirement that $\psi_{\rm cl}\rightarrow 0$ as
$t\rightarrow -i\infty$. This is a very convenient characterization of
vacuum.
As a final step in elucidating the fermionic path integral,
let us Euclideanize the classical field and the action.
To obtain expressions that resemble those found previously
for bosons, we
require that the Euclidean classical field is real; that is,
that it
is a sum of a holomorphic function and an antiholomorphic one.
We then have, finally,
$$
\Psi[\psi] = e^{-{1\over 2\pi} \int\psi{\bar\partial}\psi }~.
\eqn\finalexp
$$
It is easy to verify that the exponent is
real.
In this form,
conformal invariance is manifest.

To
exemplify the use of
this machinery, let us
calculate the vacuum wave functional.
We
proceed as for bosons: write the classical field
$$
\psi_{\rm cl} = \int_0^\infty
{dk\over\sqrt{2\pi}}~[\psi_k e^{kz}+{\bar\psi}_k
e^{k\bar{z}}]
\eqn\fermfield
$$
where we have imposed the reality condition, and the condition that
the field vanishes at early times.
Inserting this in the action we find
$$
A(\psi_{\rm cl})={1\over 2\pi}\int \psi_{\rm cl}{\bar\partial}\psi_{\rm cl}
= {1\over 2}\int {dk\over 2\pi} {\bar\psi}_k \psi_k
\eqn\fermact
$$
and for the wave functional
$$
\Psi[\psi] = \prod_{k>0}~e^{-{1\over 2}{\bar\psi}_k\psi_k}~.
\eqn\fermwf
$$
Thus this formalism indeed yields the expected
$\langle 0|\psi\rangle$.

\chapter{Geometric Entropy of Free Bosons and Fermions in 2 Dimensions}

\section{Strategy}

We now return to the problem that motivated the preceding ordeal.
We will
use the formalism developed to carry out the calculation of the
geometric entropy for
massless scalar bosons and spin ${1\over 2}$ fermions in
1+1 dimensions.  In this context, the essential problem is the
entropy associated with a half-line.
We will demonstrate that each step in
the calculation of the entropy can
be carried out very explicitly,
using the flexibility of the path integral expressions
just developed.
This may not be the
most efficient way to reach that specific goal,
but it provides an explicit, and hopefully transparent,
derivation of the
entropy in a manner that is parallel for fermions and bosons,
and capable of generalization.

The first step is to calculate the vacuum wave functional
$$
\Psi[R,L] \propto \int {\cal D}\phi~ e^{-A(\phi)}~.
\eqn\psia
$$
Here the path integral is over all fields that satisfy
$$
\phi(\sigma) = \theta(\sigma) R(\sigma) + \theta(-\sigma) L(\sigma)
\eqn\lrsplit
$$
at a time slice taken to be $\tau=0$, and we also impose
$$
\phi(x)\rightarrow 0 ~~~~\tau\rightarrow -\infty
$$
to project onto the vacuum.
In earlier sections we worked with the Fourier components of the
field but these are not localized to be either on the
left hand side of the axis, or on the right hand side.
Hence we need to choose as a basis
instead functions that are partly localized,
but still resemble Fourier modes sufficiently to
approximately diagonalize
the action.
Wavelets [\wavelet] are designed for exactly this purpose,
that is to provide wave-like functions
with compact support.  They diagonalize the action approximately,
and are likely to be of considerable use in problems with more
complicated structure\foot{There is also a more fundamental
point that ought to be mentioned in this context.  The geometric
entropy as defined corresponds to the density matrix for a hypothetical
experimenter who has complete access to arbitrarily high frequency
modes on the outside, but no access to the inside.  A more realistic
idealization would be to allow access to low-frequency modes on the
outside only, tracing over both very high frequencies in general
and also low frequencies on the inside.  These notions could be
formalized using wavelets.  The entropy thus defined would be finite,
diverging only as the limiting frequency is taken to infinity.}.
For the present however,
we will stick to Fourier modes, and simply transform the argument
of the fields
$R$ and $L$ instead.
More precisely, we will
introduce a convenient coordinate system that maps both half
lines to full lines, for which we can use the standard Fourier transform.
The density matrix will not quite diagonalize, but will break up into
2$\times$2 blocks.
This trick is in essence due to Unruh [\unruh].
The boundary conditions have a unique solution among fields
that are the sum
of a holomorphic and an anti-holomorphic piece.
That field is the classical field and the path integral is calculated
by finding the action of the classical field.  This is true, as we have
seen,
for bosons and fermions alike.

Having obtained the wave functional in a convenient
basis, the next step is to sum over the left
variables, and then
to find the entropy corresponding to the resulting density
matrix.  This will be done using a replica trick, as in
[\ccfw ].

We will use the complexification $z=\sigma+i\tau$ and
${\bar z}=\sigma-i\tau$. Since
$\tau$ is the Euclidean time this amounts simply to
the light cone coordinates.
This convention interchanges $\tau$ and $\sigma$ compared to
the one conventional in the string theory literature.

\section{Classical Fields}

First we calculate the classical field for bosons.
$\phi(x)$ is specified at $\tau=0$; our task is to determine
$\phi(x)$ in the entire lower half plane. We write
$$
\phi(z,\bar{z}) = {i\over 2\pi}\int^{\infty}_{-\infty} dw
( {1\over w - z} - {1\over w  - \bar{z}})
\phi(w)~.
\eqn\poiss
$$
In the integral $w$ is a real variable.
This integral equation is clearly
valid on the real line, and extends by regularity
to the
entire negative half-plane (it is just the usual Poisson integral
for this problem).
We change of variables according to
$$
w = {\rm sign}(w)~e^{x},~~~z=e^{\eta},~~~{\bar z}=e^{\bar\eta}~,
\eqn\chvar
$$
leaving the field untouched.
The field is defined in the lower half plane,  so in
inverting $z=e^{\eta}$
we must choose ${\rm Im}\eta\leq 0$. Thus the positive half-axis
is mapped to the entire real axis, and the negative half-axis is
mapped to the line with imaginary part $-i\pi$.
We write
$$
R(x)=\int {d\omega\over\sqrt{4\pi |\omega|}} e^{-i\omega x} r_{\omega} ~~~~
L(x)=\int {d\omega\over\sqrt{4\pi |\omega|}} e^{-i\omega x} l_{\omega}~,
\eqn\RL
$$
thus parametrizing
the functions $R$ and $L$ by their Fourier components
in the transformed variable. The reality condition on
the field $\phi$ is
expressed as $r_{\omega}={\bar r}_{-\omega}$ and
$l_{\omega}={\bar l}_{-\omega}$. Now
we have
$$
\phi(\eta)={i\over 2\pi}
\int {d\omega\over\sqrt{4\pi |\omega|}} \int^\infty_{-\infty} dx
[( {e^{-i\omega x}\over 1-e^{\eta-x} }r_{\omega}
-{e^{-i\omega x}\over 1+e^{\eta-x} }l_\omega ) - {\rm h.c.}]~,
$$
and calculating the integrals over $x$ by
contour integration and recalling ${\rm im}\eta\leq 0$ we
find
$$
\phi(\eta)=\int {d\omega\over\sqrt{4\pi |\omega|}} [
e^{-i\omega\eta}{1\over 2{\rm sh}\pi\omega}(e^{\pi\omega}r_\omega -l_\omega) -
e^{-i\omega\bar{\eta}}{1\over 2{\rm sh}\pi\omega}
(e^{-\pi\omega}r_\omega -l_\omega) ]~.
\eqn\vole
$$
It is easy check that indeed \lrsplit~is satisfied, i.e.
that $\phi=R$ for $\eta\in{\cal R}$ and $\phi=L$ for $\eta\in{\cal R}-i\pi$.
Indeed, this
expression could easily have been found by writing the general
form of the wave function
and determining the coefficients from the boundary
conditions.
In this reasoning, the boundary condition that $\phi\rightarrow 0$ at
early times, that is  the choice of vacuum, is expressed by imposing
regularity throughout the strip.
The present, constructive approach has the advantage that it is easily
generalized to the case of fermions.

Indeed, let us write
$$
\psi(z,\bar{z}) = {i\over 2\pi}\int^{\infty}_{-\infty} dw
( {1\over w - z} - {1\over w  - \bar{z} })
\psi(w)~.
\eqn\fermsplit
$$
Let us again introduce the left/right split \lrsplit , and
the change of variables
\chvar . Now, however, we transform the fermion field according to
$$
\psi(z,{\bar z}) =
e^{-{1\over 2}\eta}\psi (\eta ,{\bar\eta})
,~~~~~z=e^{\eta},~{\bar z}=e^{\bar\eta}~.
\eqn\fermtrans
$$
The necessity to transform $\psi$, in contrast to $\phi$, ultimately
reflects the non-trivial conformal weight of $\psi$.
The original $\psi$ was real and that
transforms to a real $\psi$ on the real axis but to a purely
imaginary $\psi$ on ${\cal R}-i\pi$.
Introducing Fourier transforms
$$
R(x) = \int {d\omega\over\sqrt{2\pi}} e^{-i\omega x}r_{\omega},~~~
L(x) = \int {d\omega\over\sqrt{2\pi}} e^{-i\omega x}l_{\omega}~,
$$
this is expressed by ${\bar r_\omega}=r_{-\omega}$ and
${\bar l_\omega}=-l_{-\omega}$. We have chosen
a different normalization here
than in \RL\ for bosons, in order
that the Fourier components $r_\omega$ and
$l_\omega$ have mass dimension $-{1\over 2}$ for fermions,
as it did bosons.
Collecting formulae, we have
$$
\psi(\eta,{\bar\eta})e^{-{1\over 2}\eta}
=i\int {d\omega\over\sqrt{2\pi}} \int^\infty_{-\infty} {dx\over 2\pi}
e^{-i\omega x}e^{{1\over 2}x}[(
{r_{\omega}\over e^{x}-e^{\eta} }
-{il_\omega
\over e^{x}+e^{\eta} }) - (\eta\rightarrow {\bar\eta})]~.
$$
The extra factors of $e^{{1\over 2}x}$ and $e^{{1\over 2}\eta}$
compared to the boson case come from the transformation of the fermion
fields as a $({1\over 2},0)$ field. This is also the origin of the
extra $i$ in the second term.
Upon performing the integrals
we find
$$
\psi(\eta,{\bar\eta})=\int {d\omega\over\sqrt{2\pi}} [
e^{-i\omega\eta}{1\over 2{\rm ch}\pi\omega}(e^{\pi\omega}r_\omega
+ l_\omega) + e^{\eta-{\bar\eta}\over 2}
e^{-i\omega\bar{\eta}}{1\over 2{\rm ch}\pi\omega}
(e^{-\pi\omega}r_\omega -l_\omega) ]
\eqn\fermdecomp
$$
In principle  this
decomposition could have been found by writing the {\it ansatz},
$$
\psi(\eta,{\bar\eta}) = \int {d\omega\over\sqrt{2\pi}}
 [e^{-i\omega\eta}\psi_\omega + e^{\eta-{\bar\eta}\over 2}
e^{-i\omega{\bar\eta}}{\bar\psi_\omega} ]
$$
and determining the coefficients from the boundary conditions.
However this {\it ansatz\/} is non-trivial, the $e^{\eta-{\bar\eta}\over 2}$
being due to the conformal dimensions of the fields.
In the present, constructive approach it is well motivated from the
transformation properties of the fermion field.

\section{Wave Functionals}

Having found the classical field for both bosons and fermions
we proceed to find the wave function. The classical actions
are
$$
A^{\rm boson}_{\rm cl}=
{1\over 2\pi}\int \partial\phi_{\rm cl}{\bar\partial}\phi_{\rm cl}
= {1\over 2}\int_0^\infty {d\omega\over 2\pi}
[{{\rm ch}\pi\omega\over{\rm sh}\pi\omega}({\bar r_\omega}r_\omega
+{\bar l_\omega}l_\omega) - {1\over{\rm sh}\pi\omega}
({\bar r_\omega}l_\omega+{\bar l_{\omega}}r_\omega)]
$$
and
$$
A^{\rm fermion}_{\rm cl}={i\over 2\pi}\int \psi_{\rm cl}
{\bar\partial}\psi_{\rm cl}
= {1\over 2} \int_0^\infty {d\omega\over 2\pi}
[{{\rm sh}\pi\omega\over{\rm ch}\pi\omega}(r_{-\omega}r_\omega
+l_{-\omega}l_\omega) + {1\over{\rm ch}\pi\omega}
(r_{-\omega}l_\omega-l_{-\omega}r_\omega)]
$$
In the fermion case we avoid the ${\bar l_\omega}$ notation to
prevent confusion due to the relation ${\bar l_\omega}=-l_{-\omega}$.
In evaluating these expressions we have used the integrals
$$
\int\partial e^{-i\omega^\prime \eta}{\bar\partial}e^{-i\omega{\bar\eta}}
= 2\pi\delta(\omega+\omega^\prime)\omega e^{\pi\omega}{\rm sh}\pi\omega
$$
$$
\int e^{-i\omega^\prime \eta} {\bar\partial}e^{\eta-{\bar\eta}\over 2}
e^{-i\omega {\bar\eta}} = 2\pi\delta(\omega+\omega^\prime)
e^{\pi\omega} i{\rm ch}\pi\omega
$$
$$
\int e^{\eta-{\bar\eta}\over 2}e^{-i\omega^\prime {\bar\eta}}{\bar\partial}
e^{\eta-{\bar\eta}\over 2}e^{-i\omega {\bar\eta}}=0~.
$$

The wave functionals are simply
$$
\Psi[R,L]=e^{-A_{\rm cl}}
$$
It is a good check, to verify
some of their necessary qualitative properties.
In the limit $\omega\rightarrow\infty$ the wave functionals reduce to
the results found with no left/right
split, that is, Gaussians with the same normalization.
This occurs because excitations that are almost localized do not mix
with excitations on the other half line.
At finite frequencies, there is
overlap.  We note however that the operator in the exponent has
the same determinant as previously, as indeed it
must for a unitary change of
basis.  This feature is easily checked using
the path integral measure $\prod_\omega dl_{-\omega}dl_\omega
dr_{-\omega} dr_\omega$, and the useful integral formula
$$
\int d{\bar z}dz~e^{-{\bar z}Mz+{\bar z}j+{\bar j}z}=
({\rm det}M)^{\mp{1\over 2}}~e^{{\bar j}M^{-1}j}
$$
where the upper and lower signs refer to bosons and fermions
respectively.

\section{Geometric Entropy}

To calculate the entropy from the wave functional we must
first
calculate the density matrix
$$
\rho[R,R^{\prime}] = \int {\cal D}L~{\bar\Psi}[R,L]
\Psi[R^{\prime},L]~.
$$
The measure in the integral is simply $\prod_{\omega}dl_{-\omega}dl_{\omega}$,
which we have normalized so that a Gaussian gives unity.
The wave functional $\Psi$ is
normalized by requiring ${\rm Tr}\rho = 1$ with respect to the same measure.
We find
$$
\rho[R,R^{\prime}] = \prod_{\omega>0}
{{\rm sh}\pi\omega\over {\rm ch}\pi\omega}
{\rm exp}\{ - {1\over 2{\rm sh}2\pi\omega}
[{\rm ch}2\pi\omega (|r_{\omega}|^2+|r^{\prime}_{\omega}|^2)-
(\bar r_{\omega} r^\prime_\omega+\bar r^\prime_\omega r_\omega)]\}
\eqn\densmatr
$$
for fermions as well as for bosons.
We have defined
$|r_{\omega}|^2=r_{-\omega}r_{\omega}$, and use again
$r_{-\omega}\equiv {\bar r_{\omega}}$.
In verifying this expression
it is important
to recall that integrals over Grassmann variables gives a determinant
in the numerator rather than in the denominator.

Next we want to use the
replica trick
$$
S_{\rm geom}= - (1-{d\over dn}){\rm ln}{\rm Tr}{\rho^n}~,
\eqn\replica
$$
so we need to calculate $\rho^n$. The result is
$$
{\rho^n[R,R^\prime]\over {\rm Tr}\rho^n}
= \prod_{\omega>0}
{{\rm sh}n\pi\omega\over {\rm ch}n\pi\omega}
{\rm exp}\{ - {1\over 2{\rm sh}2\pi n\omega}
[{\rm ch}2\pi n\omega (|r_{\omega}|^2+|r^{\prime}_{\omega}|^2)-
(\bar r_{\omega} r^\prime_\omega+\bar r^\prime_\omega r_\omega)]\}
\eqn\denspower
$$
where
$$
{\rm Tr} \rho^n = {(2{\rm sh}\pi\omega )^{2n}\over
(2{\rm sh}n\pi\omega )^2 }~~~~~({\rm bosons}) $$ $$
{\rm Tr} \rho^n = {(2{\rm ch}n\pi\omega )^{2}\over
(2{\rm ch}\pi\omega )^{2n} }~~~~~({\rm fermions})
\eqn\powertrace
$$
These formulae can be verified inductively.
The difference between bosons and fermions is two-fold.
First: Grassmann integrals, as mentioned above, give determinants
in the numerator rather than in the denominator.  Second:
in taking the trace for bosons we simply
identify $r_\omega=r_\omega^\prime$ and do the
$dr_{-\omega}dr_\omega$ integral,
but for fermions we must take $r_\omega=-r_\omega^\prime$
instead.
This difference has been explained in an elementary way by Soper
[\soper].
We can also understand it simply in our framework, as follows.
In the holomorphic formalism the typical bilinear operator can be
expanded as a string of variables in the form anti-holomorphic,
holomorphic, anti-holomorphic,
{\it etc}.  In taking the trace, however,
we pair the last holomorphic variable
in a string like this with the first anti-holomorphic variable.
For Grassmann variables this operation
must be accompanied by a change of sign, which is most easily handled
by changing the sign on one of the variables.  This accounts
for the antisymmetric boundary conditions.

With these expressions
the replica trick \replica\ can be carried through to yield
$$
S_{\rm geom}
= \pm 2\fint (1-\omega{d\omega\over d\omega}){\rm ln}(1\mp e^{-2\pi\omega})
= 4\int d\omega {\omega\over e^{2\pi\omega}\mp 1}
\eqn\nearfinalent
$$
for the entropy. The upper sign refers to boson and the lower to fermions.
These are simply the thermodynamic expressions for the entropy
of a 1--dimensional gas of (spinless) bosons or fermions respectively.
So far, the possible values of $\omega$ have been specified informally
as $\fint$,
and strictly speaking expressions such as $\prod_\omega$ did
not make sense without regularization. At this point  we
impose periodic (or anti-periodic) boundary conditions in a box
of length $L$, thus arriving at
$$
S_{\rm geom} = {c\over 6}L
\eqn\finalent
$$
with $c={1\over 2},1$ for fermions and bosons, respectively.
In this formula the length $L$ is the length as measured in the
{\it transformed\/}
coordinate system.
Transforming back to the original coordinates and introducing
an ultraviolet cutoff $\epsilon$ and an infrared cutoff $\Sigma$ we
can write $L= \ln {\Sigma\over\epsilon}$.
This result and
the interpretation
of the ensuing divergence have been discussed extensively from
another point of view in [\hlw].

\bigskip

\bigskip

We have now accomplished the technical task we set ourselves, to
obtain a path integral sufficiently flexible to allow us
to calculate wave
functionals and geometric entropy in a straightforward manner,
applicable
both to bosons and to fermions.
We were surprised, that to do so we had to forge
some new tools.

\bigskip

\ack{We wish to thank Curt Callan, who was involved in the genesis
of this work, for many helpful discussions.}

\endpage

\refout

\bye